\documentclass[prc,twocolumn,superscriptaddress,twoside,floatfix]{revtex4-2}

\usepackage{amssymb,epsfig}
\usepackage{xcolor}

\usepackage{graphicx}
\usepackage[outdir=./]{epstopdf}

\begin{document}

\title{Population of tetraneutron continuum in reactions of $^{8}$He on
deuterium}

\author{I.A.~Muzalevskii}
\email{muzalevsky@jinr.ru}
\affiliation{Flerov Laboratory of Nuclear Reactions, JINR,  141980 Dubna,
Russia}

\author{N.B.~Shulgina}
\affiliation{National Research Centre ``Kurchatov Institute'', Kurchatov sq.\ 1,
123182 Moscow, Russia}
\affiliation{Bogoliubov Laboratory of Theoretical Physics, JINR, 141980 Dubna,
Russia}

\author{A.A.~Bezbakh}
\affiliation{Flerov Laboratory of Nuclear Reactions, JINR,  141980 Dubna,
Russia}
\affiliation{Institute of Physics, Silesian University in Opava, 74601 Opava,
Czech Republic}

\author{V.~Chudoba}
\affiliation{Flerov Laboratory of Nuclear Reactions, JINR,  141980 Dubna,
Russia}
\affiliation{Institute of Physics, Silesian University in Opava, 74601  Opava,
Czech Republic}

\author{S.A.~Krupko}
\affiliation{Flerov Laboratory of Nuclear Reactions, JINR,  141980 Dubna,
Russia}

\author{S.G.~Belogurov}
\affiliation{Flerov Laboratory of Nuclear Reactions, JINR,  141980 Dubna,
Russia}
\affiliation{National Research Nuclear University ``MEPhI'', 115409 Moscow,
Russia}

\author{D.~Biare}
\affiliation{Flerov Laboratory of Nuclear Reactions, JINR,  141980 Dubna,
Russia}

\author{I.A.~Egorova}
\affiliation{Flerov Laboratory of Nuclear Reactions, JINR,  141980 Dubna,
Russia}

\author{A.S.~Fomichev}
\affiliation{Flerov Laboratory of Nuclear Reactions, JINR,  141980 Dubna,
Russia}
\affiliation{Dubna State University, 141982 Dubna, Russia}

\author{E.M.~Gazeeva}
\affiliation{Flerov Laboratory of Nuclear Reactions, JINR,  141980 Dubna,
Russia}

\author{A.V.~Gorshkov}
\affiliation{Flerov Laboratory of Nuclear Reactions, JINR,  141980 Dubna,
Russia}

\author{L.V.~Grigorenko}
\affiliation{Flerov Laboratory of Nuclear Reactions, JINR,  141980 Dubna,
Russia}
\affiliation{National Research Nuclear University ``MEPhI'', 115409 Moscow,
Russia}
\affiliation{National Research Centre ``Kurchatov Institute'', Kurchatov sq.\ 1,
123182 Moscow, Russia}

\author{G.~Kaminski}
\affiliation{Flerov Laboratory of Nuclear Reactions, JINR,  141980 Dubna,
Russia}
\affiliation{Heavy Ion Laboratory, University of Warsaw, 02-093 Warsaw, Poland}

\author{M.~Khirk}
\affiliation{Skobeltsyn Institute of Nuclear Physics, Moscow State University,
119991 Moscow, Russia}
\affiliation{Flerov Laboratory of Nuclear Reactions, JINR,  141980 Dubna,
Russia}

\author{O.~Kiselev}
\affiliation{GSI Helmholtzzentrum f\"ur Schwerionenforschung GmbH, 64291
Darmstadt, Germany}

\author{D.A.~Kostyleva}
\affiliation{GSI Helmholtzzentrum f\"ur Schwerionenforschung GmbH, 64291
Darmstadt, Germany}
\affiliation{II. Physikalisches Institut, Justus-Liebig-Universit\"at, 35392
Giessen, Germany}

\author{M.Yu.~Kozlov}
\affiliation{Laboratory of Information Technologies, JINR,  141980 Dubna,
Russia}

\author{B. Mauyey}
\affiliation{Flerov Laboratory of Nuclear Reactions, JINR,  141980 Dubna,
Russia}
\affiliation{Institute of Nuclear Physics, 050032 Almaty, Kazakhstan}

\author{I.~Mukha}
\affiliation{GSI Helmholtzzentrum f\"ur Schwerionenforschung GmbH, 64291
Darmstadt, Germany}

\author{E.Yu.~Nikolskii}
\affiliation{National Research Centre ``Kurchatov Institute'', Kurchatov sq.\ 1,
123182 Moscow, Russia}
\affiliation{Flerov Laboratory of Nuclear Reactions, JINR,  141980 Dubna,
Russia}

\author{Yu.L.~Parfenova}
\affiliation{Flerov Laboratory of Nuclear Reactions, JINR,  141980 Dubna,
Russia}

\author{W.~Piatek}
\affiliation{Flerov Laboratory of Nuclear Reactions, JINR,  141980 Dubna,
Russia}
\affiliation{Heavy Ion Laboratory, University of Warsaw, 02-093 Warsaw, Poland}

\author{A.M.~Quynh}
\affiliation{Flerov Laboratory of Nuclear Reactions, JINR,  141980 Dubna,
Russia}
\affiliation{Nuclear Research Institute, 670000 Dalat, Vietnam}


\author{A.~Serikov}
\affiliation{Flerov Laboratory of Nuclear Reactions, JINR,  141980 Dubna,
Russia}

\author{S.I.~Sidorchuk}
\affiliation{Flerov Laboratory of Nuclear Reactions, JINR,  141980 Dubna,
Russia}

\author{P.G.~Sharov}
\affiliation{Flerov Laboratory of Nuclear Reactions, JINR,  141980 Dubna,
Russia}
\affiliation{Institute of Physics, Silesian University in Opava, 74601 Opava,
Czech Republic}

\author{R.S.~Slepnev}
\affiliation{Flerov Laboratory of Nuclear Reactions, JINR,  141980 Dubna,
Russia}

\author{S.V.~Stepantsov}
\affiliation{Flerov Laboratory of Nuclear Reactions, JINR,  141980 Dubna,
Russia}

\author{A.~Swiercz}
\affiliation{Flerov Laboratory of Nuclear Reactions, JINR,  141980 Dubna,
Russia}
\affiliation{AGH University of Science and Technology, Faculty of Physics and
Applied Computer Science, 30-059 Krakow, Poland}

\author{P.~Szymkiewicz}
\affiliation{Flerov Laboratory of Nuclear Reactions, JINR,  141980 Dubna,
Russia}
\affiliation{AGH University of Science and Technology, Faculty of Physics and
Applied Computer Science, 30-059 Krakow, Poland}

\author{G.M.~Ter-Akopian}
\thanks{Deceased}
\affiliation{Flerov Laboratory of Nuclear Reactions, JINR,  141980 Dubna,
Russia}
\affiliation{Dubna State University, 141982 Dubna, Russia}

\author{B.~Zalewski}
\affiliation{Flerov Laboratory of Nuclear Reactions, JINR,  141980 Dubna,
Russia}
\affiliation{Heavy Ion Laboratory, University of Warsaw, 02-093 Warsaw, Poland}


\date{\today.}


\begin{abstract}
Search for the population of the low-energy continuum of a tetraneutron system
was performed for reactions of the $^{8}$He beam on a deuterium target. These
studies are based on the data [I.A. Muzalevskii \textit{et al.}, Phys.\
Rev.\ C \textbf{103}, 044313 (2021)], previously used for the studies of $^{7}$H
and $^{6}$H in the $^2\text{H}(^8\text{He},{^3\text{He}})^{7}$H and
$^2\text{H}(^8\text{He},{^4\text{He}})^{6}$H reactions. Evidence for a hump in
the $^4$n continuum at $3.5 \pm 0.7$ and $3.2 \pm 0.8$ MeV was observed in the
$^2$H($^8$He,$^6$Li)$^4$n and $^2$H($^8$He,$^3$He)$^7$H$\rightarrow ^3$H+$^4$n
reactions, respectively. The observed statistics is quite low (6 events and up
to 40 events) corresponding to very low cross sections of few microbarns or tens
of microbarns. The background conditions for the $^2$H($^8$He,$^6$Li)$^4$n
reaction
are shown to be good, favoring the physical nature of the observed events. The
$^2$H($^8$He,$^3$He)$^7$H$\rightarrow ^3$H+$^4$n process transforms to the
$^2$H($^8$He,$^6$Li$^{\ast})^4n$ reaction in the limit of the highest $^7$H
decay energies. The population of the low-energy region in the $^{4}$n spectrum
is found to be perfectly correlated with the population of the lowest $^{6}$Li
state in the $^{3}$He+$^{3}$H continuum with $E^*=18$ MeV. Theoretical
calculations of $^{8}$He in a five-body $\alpha$+$4n$ and of $^{4}$n in a
four-body hyperspherical models are presented. The $^{8}$He wave function is
shown to contain strong specific correlations, which may give rise to very
low-energy structures in tetraneutron continuum in extreme-peripheral reaction
scenarios.
\end{abstract}

\maketitle


\section{Introduction}
\label{sec:intro}


The search for multineutron systems is old, but still unsettled problem of the 
low-energy nuclear physics. The first ideas about the possible nuclear-stability 
of multineutron systems were expressed in Refs.\ \cite{Baz:1972a,Baz:1972} and 
since that time reexamined on various occasions. A detailed account of the 
search for \emph{bound} multineutron, both experimental and theoretical, can be 
found in the recent review \cite{Marques:2021}. In short, either none or only 
marginal experimental evidence, which was never confirmed later, was obtained 
for the bound tetraneutron. However, such attempts continue, and there is a very 
recent example \cite{Faestermann:2022}. The situation has changed with the 
recent studies of the low-energy $^{4}$n \emph{continuum} population in 
reactions with $^{8}$He, where four neutrons can be found in a 
spatially-separated neutron-halo configuration. In the 
$^{4}$He($^{8}$He,$2\alpha$) reaction \cite{Kisamori:2016} 4 events were 
observed in the low center-of-mass (c.m.) $^{4}$n energy range $0< E_T(4n) < 2$ 
MeV with the tiny cross section of $\approx 4$ nb. Even more recently, a 
statistically convincing peak (reported as ``resonance-like structure'' 
comprising around 44 events) was observed in a $^{1}$H($^{8}$He,$p\alpha$) 
experiment \cite{Duer:2022} at $E_T(4n)=2.37$ MeV with $\Gamma=1.75$ MeV. The 
reported cross section in this study is also small and estimated to be 
sub-microbarn. It should be noted that the both observations 
\cite{Kisamori:2016,Duer:2022} belong to the very specific situation of extreme 
backward (backward/forward in the case of \cite{Kisamori:2016}) kinematics of 
quasi-free scattering.

The issue of bound ``multineutron nuclei'' was scrutinized in the modern
theoretical approaches \cite{Pieper:2003,Timofeyuk:2003b,Higgins:2021}
with the same result: for fixed two-body potentials, radical
modifications of the three-body potential are required to get bound $^{4}$n.
Such modifications are considered unacceptable as they are inconsistent with
well-known nuclear structure of all other nuclides. In the theoretical studies
of the low-energy $^{4}$n continuum various indications were found for important
low-energy effect of the $^{4}$n final-state interaction (FSI). These include: 
specific trajectories of the $S$-matrix poles, enhanced ``time delay'', energy 
extrapolations of multi-neutron states confined in traps, etc., see e.g.\
\cite{Sofianos:1997,Lazauskas:2005,Gandolfi:2017,Fossez:2017,Deltuva:2018,%
Deltuva:2019,Higgins:2020}. There have been, however, two works that have 
predicted a $^{4}$n \emph{resonant state} \cite{Shirokov:2016,Li:2019}, strongly 
contradicting the other studies. It has been mentioned above that the 
illuminating summary of those studies can be found in the review 
\cite{Marques:2021}.

An approach to $^{4}$n system tractable in terms of observables (namely, the 
reactions with neutron-rich halos) was presented in Ref.\ 
\cite{Grigorenko:2004}. In this work the $^{4}$n continuum is populated from the 
``prearranged'' 4 neutron configurations in the ``atmosphere'' of a halo nucleus 
$^{8}$He. The approach predicts population at low $E_T(4n)$ energies, but this 
population is a result of a specific (very peripheral) initial state structure 
(ISS) sampling, modified by FSI and should not be considered a stand-alone 
resonance. The formation of such a low-energy continuum response is partly 
analogous to the mechanism of the ``soft dipole mode'' formation in halo nuclei: 
(i) relatively very weak FSI effects, but nevertheless, strong low-energy 
concentration of the strength function due to (ii) the peripheral character of 
the halo wave functions (WF) and (iii) the further periphery enhancement by 
electromagnetic transition operators 
\cite{Hansen:1987,Bertulani:1988,Grigorenko:2020}. Principal message of 
\cite{Grigorenko:2004} is that both halo character of $^{8}$He WF and $^{4}$n 
FSI are not yet sufficient to produce low-energy response in the $^{4}$n 
continuum, and additional assumption about peripheral nature of the reaction 
mechanism is required.

In Ref.\ \cite{Lazauskas:2023} the data of \cite{Duer:2022} were analyzed under 
assumptions similar to Ref.\ \cite{Grigorenko:2004}. However, conclusion was 
made in \cite{Lazauskas:2023} that the $^{8}$He-induced source can directly 
explain the data of \cite{Duer:2022} without any peripheral assumptions on 
reaction mechanism. So, there is a qualitative contradiction between the results 
of \cite{Grigorenko:2004} and \cite{Lazauskas:2023} based on more or less the 
same physical assumptions.

In this work we demonstrate that an evidence for the low-energy structures, 
analogous to the observations of \cite{Kisamori:2016,Duer:2022}, can be found in 
the other reactions with the $^{8}$He beam. As for the theoretical discussion, 
we revisit the $^{4}$n population from $^{8}$He in simple reaction approach of 
Ref.\ \cite{Grigorenko:2004} (sudden removal approximation) introducing, 
however, important technical improvements. We generally confirm the results of 
\cite{Grigorenko:2004}, but provide an additional insight into the reaction 
mechanism, connected with specific correlations in the $^{8}$He WF. Possible 
reasons for disagreement with the results of \cite{Lazauskas:2023} are analyzed.

The system of units $\hbar=c=1$ is used in this work.


\section{Experiment}
\label{sec:anc3-hh}


Experiments with the $^{8}$He beam impinging on the deuterium target was
performed at the ACCULINNA-2 facility (FLNR, JINR) having the search for the
$^{7}$H nuclide as a goal \cite{Bezbakh:2020,Muzalevskii:2021}. The experiment
appeared to be much ``richer'' than initially intended, providing interesting
results also on $^{6}$H \cite{Nikolskii:2022} and other auxiliary channels like
 $^{8}$Li, $^{9}$Li \cite{Nikolskii:2023,Nikolskii:2023b} and $^{5}$H, $^{5}$He,
$^{7}$He \cite{Muzalevskii:2023}. Here we present the results obtained for the
$^2$H($^8$He,$^6$Li)$^4n$ and $^2$H($^8$He,$^3$He)$^7$H$\rightarrow ^3$H+$^4$n
reactions, previously omitted. Since the experiment has been already well
presented in literature, we only briefly describe here the most relevant
details.

\begin{figure}
\centering
\includegraphics[width=0.85\linewidth]{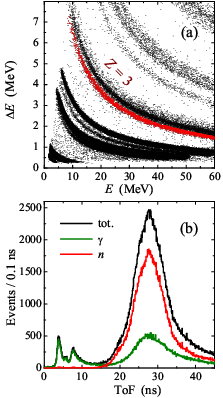}
\caption{(a) Identification of $^{6}$Li recoil nuclei (red dots) by $\Delta
E$-$E$ method in the side telescopes. (b) The ToF distribution obtained for the
stilbene-array signals triggered by the side telescopes. The set of three peaks
with ToF $<13$ ns corresponds to the gamma rays produced in the diaphragm
installed 20 cm upstream the target plane, in the target frame, and in CsI(Tl)
array. The green and red histograms are formed by the events, identified by the
$\Delta E$-TAC method as gammas and neutrons, correspondingly.}
\label{fig:ND_id}
\end{figure}

The $^{8}$He beam was produced at 26 A\,MeV with mean intensity around $10^{5}$ 
pps. The beam was focused on the cryogenic gaseous deuterium target, with a
temperature of 27\,K, equipped with the thin stainless-steel and mylar windows.
The target thickness was $3.7 \times 10 ^{20}$ $\text{atoms} \cdot
\text{cm}^{-2}$. For the $^2$H($^8$He,$^6$Li)$^4$n reaction the setup allowed us
to measure the $^{6}$Li recoil in coincidence with a neutron from the produced
unbound $^4$n system.

The beam particles were identified by the two plastic scintillators, allowing
to deduce the energy of the projectile from its measured time-of-flight (ToF).
The projectile trajectories were reconstructed by the two pairs of multi-wire
proportional chambers. A special run with empty target cell was performed
to estimate the background conditions, which was $\approx 16 \%$ of the total
$^8$He beam time long.

The recoil $^6$Li nuclei, appearing in the $^2$H($^8$He,$^6$Li) reaction hit the 
array of four identical $\Delta E$-$E$-$E$ telescopes. The telescope array was 
located 179 mm downstream the target. Each telescope consisted of three layers 
of silicon strip detectors (SSDs). The 20-$\mu$m-thick SSD with a sensitive area 
of $50 \times 50$ mm$^2$ was divided into 16 strips, the second and the third 
layers consisted of the two identical 1 mm-thick SSDs ($60 \times60$ mm$^2$ with 
16 strips). The $^6$Li recoils emitted from the deuterium gas target in the 
$^2$H($^8$He,$^6$Li) reaction were detected by this telescope array in a range 
$6^{\circ}-24^{\circ}$ in the lab system. The telescopes allowed to identify 
$^6$Li with clear separation from the other registered lithium isotopes, see 
Fig.\ \ref{fig:ND_id} (a). The central telescope was installed at the beam line 
at the distance of 323 mm behind the target. It was intended to detect tritons 
emitted with high energies at the angles smaller than $9^{\circ}$ in the lab 
system. The telescope consisted of one 1.5-mm-thick double-sided SSD ($64 \times 
64$ mm$^2$, with 32 strips on each side) followed by a square array of 16 
CsI(Tl) crystals. The crystals had a cross section of $16.5 \times 16.5$ mm$^2$ 
and thickness 50 mm each.

The group of four neutrons appears in ``free flight'' as a result of the 
$\alpha$-core removal from the $^8$He projectile. An important part of the 
$^2$H($^8$He,$^6$Li)$^{4}$n  reaction analysis was the neutron identification 
and reconstruction. The neutron-wall setup \cite{Bezbakh:2018} included 48 
stilbene scintillator crystals placed on a $0.7\times 1.1$ m$^2$ area located 2 
m downstream the target at zero angle to the beam axis. The distance between the 
50-mm thick and 80-mm diameter stilbene crystals was approximately 12 cm which 
resulted in about $30 \%$ probability for neutrons to hit a stilbene detector. 
The stilbene array provided $4.5 \%$ energy resolution and the cumulative 
single-neutron registration efficiency of $\approx 1 \%$. The probability of a 
neutron registration in coincidence with the $^{6}$Li recoil can be estimated as 
$\approx 3 \%$, taking into account that four neutrons are flying forward, 
towards the stilbene array. The $n$-gamma separation was made by means of the 
Pulse Shape Discrimination (PSD) \cite{Bezbakh:2018}. The PSD information, 
supplemented with  ToF distribution, see Fig.\ \ref{fig:ND_id} (b), leads to 
suggestion that some gamma-type signals correspond to neutron-like ToF. These 
gammas are presumably produced by neutrons interacting with the stilbene modules 
housings and should be taken as neutron events.


\subsection{$^2$H($^8$He,$^6$Li)$^{4}$n results}
\label{sec:anc3-hh-nc}

\begin{figure}
\centering
\includegraphics[width=0.47\textwidth]{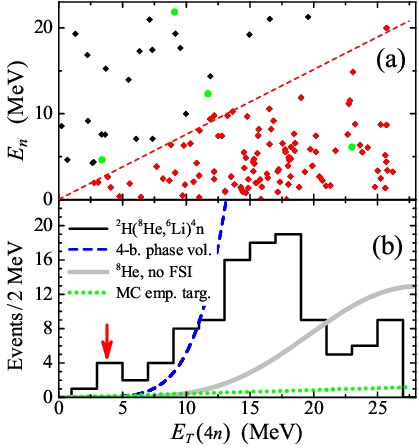}
\caption{(a) Correlation between the neutron energy in the $^{4}$n frame and the
$^{4}$n decay energy for the $^2$H($^8$He,$^6$Li)$^{4}$n reaction. The
kinematical limit $E_{n} < 3E_T(4n)/4$ is shown by
the red dashed line separating the accepted events (red diamonds) and rejected
events (black diamonds). The empty-target measurement events are shown by large
green circles. (b) The $^{4}$n MM spectrum. Green doted curve shows the
simulated empty-target spectrum assuming the uniform distribution of background
events in the $\{E_T(4n), E_{n}\}$ kinematical plane. The blue dashed curve
corresponds the 4-body
phase volume $\propto E_T^{11/2}$ and thick gray curve to Fourier transform of
the $^{8}$He source.}
\label{fig:mm_4n}
\end{figure}

Tetraneutron spectrum was reconstructed from the recoil $^6$Li as a missing-mass 
(MM) component in the $^2$H($^8$He,$^6$Li) reaction. For background reduction we 
considered $^6$Li coincidences with one of the neutrons coming from the $^{4}$n 
decay. The total number of $^6$Li-$n$ coincidences found in the recorded data 
was 136, see Fig.\ \ref{fig:mm_4n} (a). In this figure the neutron kinetic 
energy $E_n$ in the $^{4}$n c.m.\ frame is plotted opposite the reconstructed MM 
energy $E_T(4n)$ of the $^4n$ group. The fact that the majority of these events 
(108 events) are located inside the kinematically allowed region confirms a good 
channel identification for the data. The empty-target measurement Fig.\ 
\ref{fig:mm_4n} (a) shows that the background conditions were very ``clean'' for 
the $^{4}$n population in the $^2$H($^8$He,$^6$Li) reaction: the empty-target 
events are very few and located mainly in the unphysical part of the kinematical 
plane.

The MM spectrum of $^{4}$n, Fig.\ \ref{fig:mm_4n} (b), was reconstructed from
the measured $^6$Li recoil energy and emission angle, taking the 108 events
located within the ``kinematical triangle'' $E_{n} < 3 E_T(4n)/4$. The obtained
spectrum shows a group of 6 events around $E_T(4n)\approx 3.5$ MeV. The width of
this group is reasonably consistent with the value of energy resolution $\approx 
1.2 $ MeV FWHM in this region of $^{4}$n MM spectrum. The green dotted curve in 
Fig.\ \ref{fig:mm_4n} (b) shows the Monte Carlo (MC) simulated empty-target
contribution.  The statistics of the low-energy group of events is quite low, 
but the energy resolution is sufficient and the background
conditions are shown in Fig.\ \ref{fig:mm_4n} to be very good. It can be 
evaluated that less than 1 random event is expected in the energy range $0< 
E_T(4n)< 6$ MeV. Then the Poisson probability to get 6 random coincidences in
that energy region is $\lesssim 0.05 \%$.

Obviously, the observed low-energy hump in Fig.\ \ref{fig:mm_4n} (b) can not be
described just as a contribution of the 4-body phase volume or Fourier
transform of the $^{8}$He source, see also Fig.\ \ref{fig:th-comp} (d) and the
related discussion in Sec.\ \ref{sec:tetran-pop}. The phase volume for
uncorrelated decay of $A$-body system, which can be regarded here as a ``minimum
admissible model'', is
\begin{equation}
\frac{dW}{dE_T(A)} \propto E^{(3A-5)/2}_T(A) \,,
\label{eq:ph-vol-a}
\end{equation}
and the generic 4-body phase volume behavior is $\propto E_T^{7/2}$. However, 
Eq.\
(\ref{eq:ph-vol-a}) should be modified by the minimum number of excitation
quanta $n_{\min}$ allowed for the continuum WF as
\begin{equation}
\frac{dW}{dE_T(A)} \propto E^{(3A-5)/2+n_{\min}}_T(A) \,.
\label{eq:ph-vol-a-m}
\end{equation}
For the $^{4}$n system the Pauli principle requires two additional
excitation quanta for the  $0^+$ lowest-energy configuration. This corresponds
to $[s^2p^2]_0$ lowest-energy configuration in the shell-model terms or to
$K_{\min}=n_{\min}=2$ in terms of the hyperspherical harmonics (HH) model
principal quantum number $K$ (hypermomentum), see Section \ref{sec:th}. This 
leads to a modified 4-body phase-volume behavior $\propto E_T^{11/2}$ shown in 
Fig.\ \ref{fig:mm_4n}
(b).

It is notable that the mean energy $3.5 \pm 0.35\text{(stat.)}\pm 
0.6\text{(sys.)}$ MeV of the low-energy group of events in Fig.\ \ref{fig:mm_4n} 
(b) is consistent (within experimental energy resolution) with the $^{4}$n 
low-energy peak value $2.37 \pm 0.38\text{(stat.)}\pm 0.44\text{(sys.)}$ MeV 
reported in Ref.\ \cite{Duer:2022}.

The $^2$H($^8$He,$^6$Li)$^{4}$n reaction has already been used before for
tetraneutron search, but the results were not published in details, only briefly 
in conference proceedings \cite{Rich:2004,Fortier:2007}. In Ref.\ 
\cite{Rich:2004}, a resonance-like structure (with quite a low statistics) was 
reported at $E_T(4n)\approx 2.5$ MeV. In contrast, for the next analogous 
experimental run \cite{Fortier:2007} the $^{4}$n spectrum with $E_T(4n)\lesssim 
5$ MeV was found to be dominated by the carbon background due to the use of the 
CD$_2$ target. As compared to studies of \cite{Rich:2004,Fortier:2007} our 
experiment has no carbon background and additional strong background 
elimination, connected with kinematics of Fig.\ \ref{fig:mm_4n} (a), is 
possible.

Another background issue to be discussed is possibility that the
$^2$H($^8$He,$^6$Li)$^{4}$n reaction results could be contaminated by the
$^2$H($^8$He,$^6\text{Li}^*(3.56))^{4}$n reaction contribution, where
$^6\text{Li}^*(3.56)$ is the $J^{\pi}T=0^+1$ isobaric-analogue state of the
$^{6}$He(g.s.). Since the $^6\text{Li}^*(3.56)$ state decays only
electromagnetically, we cannot distinguish the latter reaction contribution
in the current experiment.

Fortunately, there is a strong argument why the population of
$^6\text{Li}^*(3.56)$ should be very small. It is well established that
reactions at typical energies of our experiment (26 $A\,\text{MeV}$) and at 
higher energies are dominated by single-step direct-reaction mechanism. Within 
this paradigm the $^{4}$n subsystem is ``spectator'' (inert during the reaction 
time)
and transfer is induced by the T-matrix between the $^{4}$He ``participant'' and
the deuterium of the target:
\[
T_{fi}=\langle ^6\text{Li(g.s.)} \vert \sum_i \nolimits V_i \vert
^{4}\text{He+}^{2}\text{H}  \rangle \,,
\]
where $\sum_i V_i$ is sum of all possible nuclear interactions in the channel of
interest. However,  it is well known that the $^6\text{Li}^*(3.56)$ $0^+1$ state
can be obtained in $^2$H capture on $\alpha$-particle only by electromagnetic M1
transition or by parity-non-conservation neutral-current weak interaction:
\begin{eqnarray}
T_{fi} & = & \langle ^6\text{Li}^*(3.56) \vert \sum_i \nolimits V_i \vert
^{4}\text{He+}^{2}\text{H}  \rangle \equiv 0 \,, \nonumber \\
T^{(\text{pv})}_{fi} & = & \langle ^6\text{Li}^*(3.56) \vert V_{\text{pv}} \vert
^{4}\text{He+}^{2}\text{H} \rangle \neq 0 \,, \nonumber
\end{eqnarray}
where the parity-violating nuclear interaction $V_{\text{pv}}$ is known to be
extremely small. Indeed, the cross section for the capture reaction
$^4$He($d$,$\gamma)^6$Li has been measured in a broad energy range by detecting
gammas or recoil $^6$Li ions \cite{Ajzenberg:1988}. No 3.56 MeV resonance
population has been observed and the best limit on the parity-forbidden width of
this state was obtained as tiny $\Gamma_{\alpha} \leq 6.5 \times 10^{-7}$ eV in
Ref.\ \cite{Robertson:1984}. Thus, the $^2$H($^8$He,$^6\text{Li}^*(3.56))^{4}$n
reaction may be connected only with two-step and more complicated reaction
scenarios, which are expected to be suppressed.

It should be also understood that actual population $^6\text{Li}^*(3.56)$ will 
not dismiss the low-energy $^4$n issue, just make its interpretation more 
complicated. Assume extreme situation that only $^6\text{Li}^*(3.56)$ was 
actually populated. This means that we miscalculate the missing mass of $^4$n, 
which should be actually 3.56 MeV smaller according to energy-conservation. So, 
the low-energy $^4$n events in Fig.\ \ref{fig:mm_4n}, which have energies $E_T$ 
between 2.5 and 5 MeV after recalculation will have energies from $-1$ to 1.5 
MeV with centroid at around the $^4$n threshold. Existence of such a ``severe 
low-energy'' $^4$n correlation, is even less realistic and complicated to 
interpret theoretically.


\subsection{$^2$H($^8$He,$^3$He)$^7$H$\rightarrow ^3$H+$^4\text{n}$ results}


\begin{figure}
\begin{center}
\includegraphics[width=0.48\textwidth]{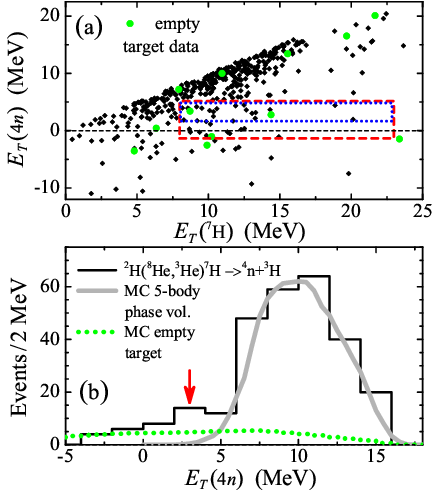}
\end{center}
\caption{(a) ``Kinematical triangle'' for $^{4}$n  MM vs.\ $^{7}$H MM for the
$^2$H($^8$He,$^3$He)$^7$H data of \cite{Muzalevskii:2021}. Large green circles
show the empty target events. Red dashed and blue dotted frames show the event
selection for Fig.\ \ref{fig:7h-t-3he} (b). (b) Tetraneutron MM spectrum summed
up for $E_T(^7\text{H}) > 8$ MeV. Green doted curve shows the
simulated empty-target spectrum assuming the uniform distribution of background
events in the $\{E_T(^7\text{H}),E_T(4n)\}$ kinematical plane. Thick gray line
is MC simulated phase volume for $^{7}$H
decays.}
\label{fig:7h-decay}
\end{figure}

In Ref.\ \cite{Muzalevskii:2021} the excitation spectrum of $^{7}$H was
populated in the $^2$H($^8$He,$^3$He)$^7$H reaction up to $E_T(^7\text{H})\sim
17$ MeV above the 5-body $^3$H+$4n$ decay threshold. An evidence was reported
for resonant states at $E_T(^7\text{H})=2.2(5)$ and 5.5(3) MeV and also some
indications for such states at 7.5(3) and 11.0(3) MeV. These structures were
observed at small c.m.\ reaction angles $\theta_{\text{c.m.}} < 18^{\circ}$. The
$^{7}$H excitations with $6< E_T(^7\text{H}) < 17$ MeV observed in a broad 
angular range $18^{\circ}< \theta_{\text{c.m.}}< 43^{\circ}$ form quite a 
featureless hump. It is possible to search for the low-energy-correlated $^{4}$n 
emission off these sufficiently highly-excited configurations of $^{7}$H.

\begin{figure}
\begin{center}
\includegraphics[width=0.49\textwidth]{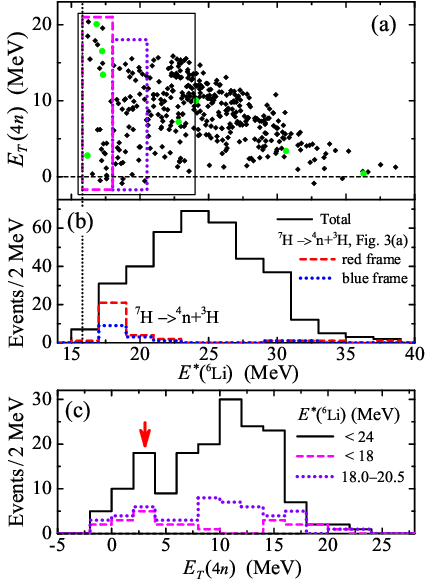}
\end{center}
\caption{(a) ``Kinematical triangle'' for $^{4}$n  MM vs.\ $E^{\ast}(^{6}$Li) MM
for the $^2$H($^8$He,$^3$He)$^7$H data of \cite{Muzalevskii:2021}. Large green
circles show the empty target events.
(b) $^{6}\text{Li}^{\ast}$ MM spectrum (black solid histogram). Red dashed and
blue dotted histograms show spectra corresponding to selection frames in Fig.\
\ref{fig:7h-decay} (a). The vertical doted line in panels (a) and (b) indicate
the $^3$He+$^3$H threshold at $E^{\ast}(^{6}\text{Li})=15.8$ MeV. (c) 
Tetraneutron MM
spectrum gated on the lowest states
in the $^{3}$He+$^{3}$H continuum of the $^{6}\text{Li}^{\ast}$ system, see
frames of corresponding style in the panel (a).}
\label{fig:7h-t-3he}
\end{figure}

The data of Ref.\ \cite{Muzalevskii:2021} contain coincidence information on the
$^{3}$H decay product of $^{7}$H, which allows one to infer the MM of the
$^{4}$n subsystem. These data in terms of the $^{4}$n emission are shown in
Fig.\ \ref{fig:7h-decay} (a). The $^{7}$H events are mainly concentrated near 
the diagonal $E_T(4n)=E_T(^7\text{H})$, which means that the ``internal'' energy
of the tetraneutron tends to be large, while the triton gets only a small
fraction of the total $^{7}$H decay energy. This situation is close to the
phase-volume distribution of the 5-body $^3$H+$4n$ decay for the given $^{7}$H
decay energy $E_T(^7\text{H})$ as
\begin{equation}
\frac{dW}{d E_T(^7\text{H}) \,d \varepsilon} \propto
E^{(\alpha+3)/2}_T(^7\text{H})\, \sqrt{\varepsilon^{\alpha}(1-\varepsilon) } \;,
\label{eq:5b-pv}
\end{equation}
in terms of $^{4}$n energy-distribution parameter
\[
\varepsilon = E_T(4n)/E_T(^7\text{H}) \; .
\]
The standard value for the 5-body phase volume is $\alpha=7$. Eq.\
(\ref{eq:5b-pv}) is the double-differential form of the Eq.\
(\ref{eq:ph-vol-a-m}), discussed in the
previous Section, so it can be found that at least $\alpha=11$ value is
requested for $\varepsilon \rightarrow 0$ by the Pauli principle for $^{4}$n.
This idea is confirmed by the
calculations of the 5-body decay of $^{7}$H in \cite{Sharov:2019}. The
$\varepsilon$ distributions at given $^{7}$H energies are also quite
``compressed'' towards small $^{3}$H energies due to the experimental bias, see 
Fig.\ 13 of \cite{Muzalevskii:2021}, so  $\alpha=11$ for the distribution of 
Eq.\ (\ref{eq:5b-pv}) turns out to be completely justified in our situation.

To observe a possible low-energy $^{4}$n correlation, we need to find kinematic 
region, where the main ``phase-volume'' component of the decay distribution Eq.\ 
(\ref{eq:5b-pv}) is well suppressed. This means $\varepsilon \lesssim 0.5$, and, 
therefore, for studies of the $^{4}$n excitation region $E_T(4n) \sim 2-4$ MeV 
we need to consider the $^{7}$H decay energy region $E_T(^7\text{H})> 8$ MeV. 
Under this selection condition, the hump with $E_T(4n) \sim 2-4$ MeV can be 
found in the $^{4}$n MM spectrum, see Fig.\ \ref{fig:7h-decay} (b). The MC 
simulated phase volume only weakly ``penetrates'' in the $E_T(4n)\sim 2-4$ MeV 
region. The background of about $2-3$ events/MeV  is estimated from the 
available empty target data, see green circles in Fig.\ \ref{fig:7h-decay} (b). 
Thus a significant part of this low-energy $^{4}$n data \textit{may be} 
connected (but not necessarily \textit{is connected}, as we see below) with a 
random background.

The $^{7}$H data can be converted to presentation of $^{6}\text{Li}^{\ast}$
system reconstructed as $^{3}$He+$^{3}$H continuum, see Fig.\ \ref{fig:7h-t-3he}
(a). Fortunately, it appeared that the low-energy $^{4}$n events in Fig.\
\ref{fig:7h-decay} (b) are practically perfectly correlated with the population
of the lowest $^{6}\text{Li}^{\ast}$ state in the $^{3}$He+$^{3}$H continuum,
with $E^{\ast}(^{6}\text{Li}) = 18$ MeV and $\Gamma \approx 3$ MeV
\cite{Tilley:2002}, see Fig.\ \ref{fig:7h-t-3he} (b). This means that we
are actually dealing here not with the $^7\text{H} \rightarrow ^3\text{H}+^4n$ 
decay, but
rather with the $^2$H($^8$He,$^6$Li$^{\ast})^4$n reaction. Moreover, the same
correlation is perfectly true not only for events in the $2< E_T(4n) <5$ MeV
range, but also for the $-1.5<E_T(4n) <2$ MeV range. In this range most of the
events should be either unphysical or connected with insufficient energy
resolution, as the population of the $E_T(4n) \lesssim 1$ MeV range is predicted
to be negligible in available theoretical scenarios, see Fig.\
\ref{fig:shulg-spec} or Ref.\ \cite{Lazauskas:2023}. The MC estimated energy
resolution for the $^{4}$n MM spectrum is around $2.5$ MeV FWHM. However, this
resolution has a broad ``nongaussian'' component, which may result in some
number of events with $E_T(4n) \sim -1$ MeV. The coincidence with the definite
state in $^{6}$Li practically guarantee the physical nature of such event, so we
recognize that the absolute majority of events (26 out of 28) found in the range
$-1.5<E_T(4n) < 5$ MeV in Fig.\ \ref{fig:7h-decay} (b) as really belonging to
the $^{4}$n MM spectrum.

Strong correlation found in Fig.\ \ref{fig:7h-t-3he} (b) inspired us to
``reverse'' the logic and construct the $^{4}$n spectrum gated by the lowest
states in the $^{6}\text{Li}^{\ast}=^{3}$He+$^{3}$H continuum, see Fig.\ 
\ref{fig:7h-t-3he} (c). The profile of the low-energy $E_T(4n) < 6$ MeV $^{4}$n 
spectrum is
practically insensitive to the $^{6}\text{Li}^{\ast}$ gate energy selection for
$E^{\ast}(^{6}\text{Li})< 24$ MeV. For higher $E^{\ast}(^{6}\text{Li})$ energies
the $^{4}$n spectrum expectedly becomes phase-volume-like. The peak energy can
be found as $3.2 \pm 0.45\text{(stat.)} \pm 0.7$(sys.) MeV. That peak position
is consistent with $2.37 \pm 0.38\text{(stat.)}\pm 0.44$(sys.) MeV of Ref.\
\cite{Duer:2022} within the experimental energy resolution. Statistically our
data is analogous to data \cite{Duer:2022}: 44 events in \cite{Duer:2022} vs.\
26 and 40 events in the $\leq 20.5$ and $\leq 24$ MeV $E^{\ast}(^{6}\text{Li})$
gates, respectively.

There is a strong sensitivity of the higher-energy part of the obtained $^{4}$n 
MM spectra to the energy of the populated $^{6}$Li states. For example, it can 
be seen in Fig.\
\ref{fig:7h-t-3he} (c) that $^{4}$n spectra in the $E^{\ast}(^{6}\text{Li})$
gates $\{15.8,18\}$ and $\{18,20.5\}$ MeV show very different populations of 4 
and 25 events in the $8 <E_T(4n) < 16$ MeV range, which seems to be beyond 
statistical uncertainty. This may be indication that mechanisms for $^{4}$n 
population within these gates are somewhat different: the states of the $^{6}$Li 
recoil are broad and overlapping in this energy range \cite{Tilley:2002}, and 
the higher-energy $E^{\ast}(^{6}\text{Li})$ gate could be ``contaminated'' by
contributions of the higher excitations of $^{6}$Li.


\section{On tetraneutron population in different reactions}
\label{sec:popul}


The low-energy population of $^{4}$n in Ref.\ \cite{Duer:2022} and in the 
processes considered in this work all have the $^{8}$He nuclide as a starting 
point. At first glance, this is the only similarity, and the reaction mechanisms 
should be too different. Let's demonstrate that the differences are actually not 
so striking.

Table \ref{tab:mom-vel} shows the momenta and velocities in the recoil-product
channel for the reaction of Ref.\ \cite{Duer:2022} and the reactions considered
in this work. Of course, in the reaction of \cite{Duer:2022} the ``recoil''
particles are leaving the FSI region faster, than in the case of our $(d,^6$Li)
reaction, but in the timescale this ``faster'' is only a factor of 2.

\begin{figure}
\begin{center}
\includegraphics[width=0.48\textwidth]{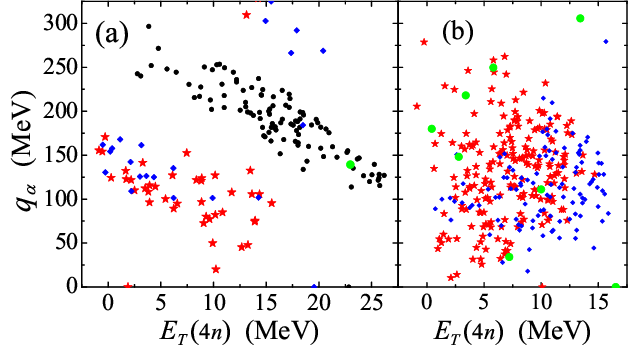}
\end{center}
\caption{Momentum transfer to $^{4}$n system for the $^2$H($^8$He,$^6$Li)$^{4}$n
reaction is shown by black circles in (a). The same for the
$^2$H($^8$He,$^3$He)$^7$H$\rightarrow\,^3$H+$^4$n reaction: blue diamonds and
red stars in (a) correspond to $E^{\ast}(^{6}\text{Li})$ ranges $\{15.8,18\}$
and  $\{18,20.5\}$ MeV,  blue diamonds and red stars in (b) correspond to
$\{20.5,24\}$ and  $\{24,40\}$ MeV. Large green circles show the corresponding
empty target events.}
\label{fig:mom-trans}
\end{figure}

\begin{table}[b]
\caption{Relative momentum $K'$ and relative velocities $v'$ in the
recoil-product channel for reactions of interest.}
\vspace{1mm}
\begin{ruledtabular}
\begin{tabular}[c]{lccc}
Reaction & Recoil & $K'$ (MeV) & $v'$  \\
\hline
$^1$H($^{8}$He,$p\, \alpha$)$^{4}$n, Ref.\ \cite{Duer:2022} & $^{5}$Li & 750 &
0.36 \\
$^2$H($^{8}$He,$^6$Li)$^{4}$n  & $^{6}$Li & 426 & 0.19 \\
$^2$H($^8$He,$^3$He)$^7$H   & $^{3}$He   & 311 & 0.16 \\
$^7$H$\,\rightarrow\,^3$H+$^4$n & $^{3}$H & $120-200$ & $0.08-0.13$ \\
\end{tabular}
\end{ruledtabular}
\label{tab:mom-vel}
\end{table}

The $^2$H($^8$He,$^3$He)$^7$H$\rightarrow ^3$H+$^4$n process formally has the 
$^7$H$\rightarrow ^3$H+$^4$n decay process as its second step. In fact, from 
Table \ref{tab:mom-vel} it is clear that for the relatively highly-excited 
states of $^{7}$H, which we consider, the velocity of the $^{3}$H recoil leaving 
$^{4}$n in the ``decay process'' is actually comparable with the velocity, at 
which the $^3$He recoil had left the FSI region at the first [the 
$^2$H($^8$He,$^3$He)$^7$H reaction] step of this process. This kinematical 
region is actually a transition region from $^2$H($^8$He,$^3$He)$^7$H reaction 
to $^2$H($^{8}$He,$^6\text{Li}^{\ast})^{4}$n reaction, where highly excited 
states of $^{6}$Li (those located above the $^{3}$He-$^{3}$H threshold) are 
populated. We find in Fig.\ \ref{fig:7h-t-3he} (c) that the most pronounced 
low-energy $^{4}$n hump, not ``contaminated'' by higher-energy phase-volume-like 
contributions, is obtained for the $^6\text{Li}^{\ast}$ recoil, which is the 
lowest-energy resonant state in the $^{3}$He-$^{3}$H continuum.

So, we see that the low-energy $^{4}$n continuum is confidently populated in the
$^2$H($^{8}$He,$^6\text{Li}^*)^{4}$n reactions with excitation energies
$E^{\ast}(^{6}\text{Li})=0$ MeV, see  Fig.\ \ref{fig:mm_4n} (b), and
$E^{\ast}(^{6}\text{Li})\approx 18$ MeV, see Fig.\ \ref{fig:7h-t-3he} (c). It is
quite natural to expect that such a population should also take place at the
intermediate energies, which correspond to $\alpha$+$d$ continuum states of
$^{6}$Li. In this case we deal with reaction of quasifree scattering 
$^2$H($^8$He,$d\,\alpha$)$^4$n, which has never been considered seriously 
because of extremely ``soft'' nature of the $^2$H target. Unfortunately we were 
not able to extract this information from our existing data due to technical 
limitations.

Another characteristic, which is expectedly very important for the direct
reactions, is momentum transfer $q_{\alpha}$ to the $^{4}$n system, see, e.g.\ 
the Eq.\ (\ref{eq:ovlp-8he}). For our reactions the reconstructed $q_{\alpha}$
values are shown in Fig.\ \ref{fig:mom-trans}. Quite different momentum 
transfers $q_{\alpha} \sim 130$ and $q_{\alpha}\sim 250$ MeV are realized for 
the low-energy $^{4}$n population in the $^2$H($^8$He,$^6$Li)$^4n$ and
$^2$H($^8$He,$^3$He)$^7$H$\rightarrow ^3$H+$^4$n reactions, respectively. This
is found important for data interpretation, see Fig.\ \ref{fig:shulg-spec} and 
related text. Unfortunately, we do not find information on the momentum 
transfers in Ref.\ \cite{Duer:2022}, though this information is obviously 
accessible in these data.
Only the restriction $0<q_{\alpha}<250$ MeV can be roughly estimated from Fig.\
2 of Ref.\ \cite{Duer:2022}, while the most probable value seem to be around
$q_{\alpha} \sim 60$ MeV.


\section{Theoretical model}
\label{sec:th}


\begin{table}[b]
\caption{Geometry of  $^{8}$He (rms radial characteristics in fm) versus
experimental data.}
\vspace{1mm}
\begin{ruledtabular}
\begin{tabular}{lcccccc}
  & $\langle r_{nn} \rangle$  & $\langle r_{\alpha n} \rangle$ & $\langle
  r_{\alpha } \rangle$ & $\langle r_{n} \rangle$ & $R_{\text{mat}}$ &
  $R_{\text{ch}}$ \\
\hline
Th. & $4.15$ & $3.34$ & $1.06$& $2.72 $ & $2.34$ & $1.96$  \\
Exp.  &  &   &  & $2.71(7) $ & $2.48(3)$ & $ 1.956(16)$  \\
\end{tabular}
\end{ruledtabular}
\label{tab:geom}
\end{table}

In this work we performed studies of $^{8}$He in a five-body $\alpha$+$4n$ and
of $^{4}$n in a four-body hyperspherical harmonics models. A detailed
description of these studies will be given elsewhere, while here we focus on the
aspects related to $^{4}$n production from $^{8}$He, following the ideology of
Ref.\ \cite{Grigorenko:2004}. The HH Schr\"{o}dinger equation (SE) for $A$
particles is used either without (bound states) or with right-hand-side
inhomogenity (continuum states):
\begin{equation}
\left(\hat{H}_A-E_T\right)\Psi_A^{(b,+)}=F_{\mathbf{q}} \,.
\label{eq:shred}
\end{equation}
On the properly antisymmetrized HH basis ${\cal J}_{K \gamma}(\Omega_{\rho})$
the variational procedure reduces the SE  with inhomogeneous term
(\ref{eq:shred}) to a set of coupled ordinary differential equations:
\begin{eqnarray}
\Psi_A^{(b,+)}(\rho,\Omega_{\rho}) = \rho^{-N_A} \sum \nolimits _{K\gamma}
\chi^{(b,+)}_{K \gamma}(\rho) {\cal J}^{\dagger}_{K\gamma}(\Omega_{\rho}) \,,
\nonumber \\
\left[  \frac{d^{2}}{d\rho^{2}}-\frac{{\cal L}({\cal L}+1)}{\rho^{2}}+
2M \left\{  E_T-V_{K \gamma,K \gamma}(\rho) \right \} \right]
\chi_{K \gamma} (\rho) \nonumber \\
= \sum \nolimits_{K' \gamma '}2M \, V_{K \gamma,K^{ \prime
}\gamma^{\prime}}(\rho) \, \chi_{K^{\prime}
\gamma^{ \prime }}(\rho) + f_{\mathbf{q},K \gamma} (\rho) \,,
 \;
 \label{eq:shred-part} \\
%
V_{K\gamma,K^{\prime}\gamma^{\prime}}(\rho)=\int d\Omega_{\rho} \;
{\cal J}_{K \gamma}^{\dagger}(\Omega_{\rho})\Big[  V_{2} + V_{3}
\Big]  {\cal J}_{K^{\prime}\gamma^{\prime}}(\Omega_{\rho})\, ,
\label{eq:v-hh} \\
V_{2}=\sum \nolimits_{i>j}\hat{V}({\bf r}_{ij})\, ,\quad V_{3}=\sum
\nolimits_{i>j>k}\hat{V}({\bf r}_{ij},{\bf r}_{jk}) \, , \nonumber
%
%
%
%
%
\end{eqnarray}
where $\hat{V}({\bf r}_{ij})$ are two-body potentials $V_{2N}({\bf r}_{ij})$ or
$V_{\text{Core-}N}({\bf r}_{ij})$, while $\hat{V}({\bf r}_{ij},{\bf r}_{jk})$
are three-body potentials $V_{3N}({\bf r}_{ij},{\bf r}_{jk})$ or
$V_{\text{Core-}NN}({\bf r}_{ij},{\bf r}_{jk})$.  The ``multiindex''
$\{K\gamma\}$ is numbering the available HH basis states. The hypermomentum $K$
is the grand quantum number in the few-body configuration space and
\[
{\cal L}=K+(3A-6)/2 \, , \qquad N_A=(3A-4)/2 \,.
\]
For $^{8}$He and $^{4}$n its minimum value is $K_{\min}=2$ due to the
Pauli principle for four neutrons.

For the SE (\ref{eq:shred-part}) without Coulomb interaction the boundary
conditions at large $\rho$ are known analytically
\begin{eqnarray}
\chi^{(b)}_{K \gamma}(\rho) & \propto & \sqrt{2 \varkappa \rho/\pi } \, {\cal
K}_{{\cal L}+1/2} (\varkappa \rho) \propto \exp[- \varkappa \rho] \,, \nonumber 
\\
\chi^{(+)}_{K \gamma}(\rho) & \propto & {\cal H}^{(+)}_{\cal L} (\varkappa \rho)
\propto \exp[+i \varkappa \rho] \,,
\label{eq:bc}
\end{eqnarray}
where $\varkappa=\sqrt{2M |E_T|}$, ${\cal K}$ are the modified Bessel functions
of the second kind, ${\cal H}^{(+)}$ are the Riccati-Bessel functions.

The ``source functions'' $f_{\mathbf{q},K\gamma} (\rho)$ are terms of
hyperspherical expansion of the inhomogeneous  term $F_{\mathbf{q}}$:
\begin{equation}
f_{\mathbf{q},K\gamma} (\rho) = \rho^{N_A} \int d \Omega_{\rho} \,{\cal J}
^{\dagger} _{K\gamma}(\Omega_{\rho})\, F_{\mathbf{q}} (\rho, \Omega_{\rho}) \,.
\label{eq:sour-def}
\end{equation}
The simplest reaction model which can be considered for $^{4}$n production  is
``sudden removal'' of core from $^{8}$He \cite{Grigorenko:2004,Lazauskas:2023}.
This procedure defines the source function $F_{\mathbf{q}}$ in (\ref{eq:shred})
as the Fourier transform of the overlap integral of the $\alpha$-cluster WF
$\Psi_{\alpha}$ and the $^8$He WF over the radius-vector $\mathbf{r}_{\alpha}$
between the removed $\alpha$-cluster and the $^4$n center-of-mass:
\begin{equation}
F_{\mathbf{q}_{\alpha}} =
\int d^3r_{\alpha} \, e^{i \mathbf{q}_{\alpha} \mathbf{r}_{\alpha}} \langle
\Psi_{\alpha} | \Psi_{^8\text{He}} \rangle \, .
\label{eq:ovlp-8he}
\end{equation}
Within the sudden removal approximation the momentum transferred to the $^{4}$n 
system is equal to the momentum $\mathbf{q}_{\alpha}$ of the removed 
$\alpha$-cluster. The ``strength function'' for population of the $^{4}$n 
continuum should be proportional to the outgoing flux of $A=4$ particles, 
expressed via the WFs with pure outgoing wave boundary conditions Eq.\ 
(\ref{eq:bc})
\begin{equation}
\frac{dW}{dE_T} \sim j=\frac{1}{M} \mathop{\rm Im} \!
\int \! d\Omega _{\rho } \left. \Psi_A ^{(+)\dagger}
\rho^{N_A}\frac{d}{d\rho} \, \rho^{N_A} \,
\Psi_A^{(+)}\right| _{\rho =\rho_{\max} } \, ,
\label{eq:current}
\end{equation}
on a hypersphere of a large radius $\rho_{\max} \sim 300-500$ fm.


\section{Properties of the $^{8}$He}


Following the cluster $\alpha$+$n$+$n$ model studies of $^{6}$He, $^{6}$Li,
$^{6}$Be isobar \cite{Zhukov:1993,Grigorenko:2009c,Grigorenko:2020} we use the
Sack-Biedenharn-Breit (SBB) potential  \cite{Sack:1954} in the $\alpha$-$n$
channel. We suppress the Pauli forbidden state by using the additional repulsive
core in $s$-wave potential, which provides practically the same scattering
phases as the deep potential.

\begin{table}[b]
\caption{Elastic and reaction cross sections for $^{8}$He (in mb).}
\vspace{1mm}

\begin{ruledtabular}
\begin{tabular}{lccc}
Reaction & $^{8}$He-$p$, $\sigma_{el}$ & $^{8}$He-$p$, $\sigma_{r}$ &
$^{8}$He-$^{12}$C, $\sigma_{r}$ \\
\hline
Th. & $51.6$ & $196.6$ & $807$ \\
Exp. & $54.1(17)$  \cite{Neumaier:2002} & $197.8(35)$  \cite{Neumaier:2002} &
$817(6) $ \cite{Tanihata:1985a} \\
\end{tabular}
\end{ruledtabular}
\label{tab:8He-cs}
\end{table}

In the calculations we adopted the local charge-independent Argonne SSC AV14
$NN$ potential \cite{Wiringa:1995}. For evaluations we also used simple $s$-wave
singlet $NN$ potential Brown-Jackson (BJ) from \cite{Brown:1976}: $V_{2N}(r)=V_0
\,
\exp[-r/r_0]$, $V_0=-31$ MeV, $r_0=1.8$ fm. For three nucleons with separations
 $r_{ij}$ we use the three-nucleon potential \cite{Hiyama:2004}
\begin{equation}
V_{3N}=\sum \nolimits_{n=1,2} U_{n} \sum \nolimits_{i>j>k}
\exp[-\mu_{n}(r_{ij}^2+r_{jk}^2+r_{ki}^2)] \,,
\label{eq:v3n}
\end{equation}
in Eq.\ (\ref{eq:v-hh}). A set of the parameters $U_{1}=-2.04$ MeV,
$\mu_{1}=(4.0 \;\text{fm})^{-2}$, $U_{2}=35.0$ MeV, $\mu_{2}=(0.75
\;\text{fm})^{-2}$ yields the binding energies 8.41 (8.48), 7.74 (7.72), and
28.44 (28.30) MeV for $^{3}$H, $^{3}$He, and $^{4}$He ground states,
respectively (the experimental values are shown in parentheses).

\begin{figure}
\begin{center}
\includegraphics[width=0.499\textwidth]{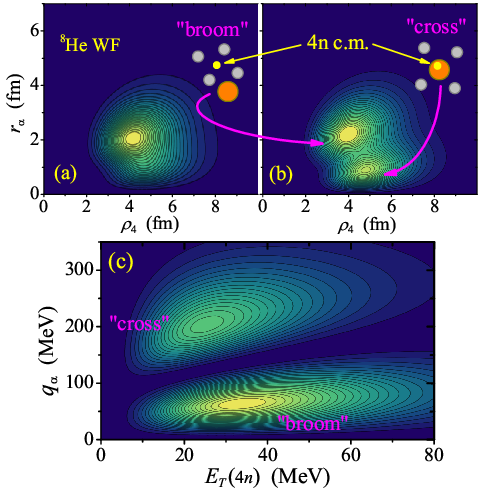}
\end{center}
\caption{(a) Correlation density of the $^{8}$He WF. (b) Correlation density of
the $0^+$ $^{4}$n source component with $L_{\alpha}=0$ with qualitative
illustrations of dominant correlation patterns. (c) Momentum-space correlations
for dominant $K=2$, $S=0$ component of the source (b).}
\label{fig:shulg-sour}
\end{figure}

We also adjusted the $^4$He-$n$-$n$ three-body potential of the same form as
Eq.\ (\ref{eq:v3n}) to reproduce the ground state properties of $^6$He, which is
well described in terms of the three-body $^4$He-$n$-$n$ model
\cite{Zhukov:1993,Grigorenko:2009c,Grigorenko:2020}. For $^{8}$He g.s.\ the
calculations were performed with $K_{\max}=10$, containing 15862 basis states
which are reduced to 675 states in (\ref{eq:shred-part})  by antisymmetrization.
The $^{8}$He is slightly underbound (by approximately 115 KeV). This means some
overbinding at higher $K$, but this does not matter for the study of the
tetraneutron: the geometry of the $^{8}$He g.s.\ WF is much more important here.
The percentages of the major $^{8}$He WF components $\{^1S,{^3P}\}=\{73,27\}$
are in a good agreement with advanced \textit{ab initio} Quantum Monte Carlo
calculations of \cite{Wiringa:2000}  $\{71,29\}$ and \cite{Pieper:2004}
$\{63,37\}$. The geometric characteristics of $^8$He are compared with
experimental data in the Table \ref{tab:geom}. The root mean square (rms) charge
radius $R_{\text{ch}}$ of $^8$He, has been determined for the first time as
1.93(3) fm \cite{Mueller:2007}. However, using the new charge radius
1.67824(83) fm of $^4$He \cite{Krauth:2021} as an anchor point for the isotope
shift measurements, the $^{8}$He charge radius was reevaluated as 1.9559(7)(158)
fm, where the first uncertainty is from  new charge radius value and the second
uncertainty from the electronic isotope shift measurements. Our result for
$^8$He charge radius is in nice agreement with both experiments. The comparison
with experimental rms matter radius $R_{\text{mat}}$ in Table \ref{tab:geom},
seems to be not so favorable, but we need to recall that this value is actually
extracted from the high-energy reaction data in a model-dependent way.
Therefore, here the direct comparison of cross sections is more preferable. The
results of Glauber-like
calculations with $^{8}$He densities obtained in our model, shown in the
Table \ref{tab:8He-cs}, are in a very good agreement with experimental data,
including the cross section on the $^{12}$C target \cite{Tanihata:1985a}, from
which the matter radius of $2.48(3)$ fm, shown in Table \ref{tab:geom} was
actually extracted. Thus, we can expect that our $^{8}$He WF, which near
perfectly describes all known experimental data and reveals extremely important
$\alpha$-$^4$n correlations is a reliable starting point for the tetraneutron
study.


\section{The $^{8}$He WF as a source for tetraneutron population}
\label{sec:tetran-pop}


The $^{8}$He WF correlation density $r_{\alpha}$ vs.\ $\rho_4$ (collective 
radius of four neutrons, informative for
the $^{4}$n studies) is shown in Fig.\ \ref{fig:shulg-sour} (a). It looks
overall quite featureless. In contrast, the projection of the $^{8}$He WF on the
$^{4}$n $0^+$ configuration with $L_{\alpha}=0$, where strongest $^{4}$n FSI
effect is expected, exhibits strong spatial correlations, see Fig.\
\ref{fig:shulg-sour} (b). The correlation density $r_{\alpha}$ vs.\ $\rho_4$ can
be easier perceived basing on a simple relation between the rams radius $r_n$ of
individual neutron in the $^{4}$n c.m.\ frame and the $^{4}$n rams hyperradius
\[
\langle r_n  \rangle =\langle \rho_4 \rangle / 2 \;.
\]
Following the well-known naming of spatial correlations in the classical
$^{6}$He halo nucleus --- ``dineutron'' and ``cigar'' \cite{Zhukov:1993}  --- we
characterize correlations in $^{8}$He source as ``broom'' (compact tetraneutron
aside of $\alpha$-particle) and ``cross'' (neutrons evenly distributed around
central-located $\alpha$-particle). The double-hump picture of correlations can
be qualitatively understood as connected with $[s^2p^2]_0$ configuration (in the
shell model notations) in analogy with $[p^2]_0$ configuration providing strong
spatial focusing (``Pauli focusing'' \cite{Zhukov:1993}) in the $^{6}$He case.
Fig.\ \ref{fig:shulg-sour} (c) shows the Fourier transform of the source (b),
illustrating the momentum content of this structure without $^{4}$n FSI. So far,
the Pauli focusing for the $^{8}$He g.s.\ WF has been several times
discussed in the qualitative models \cite{Zhukov:1994,Mei:2012,Sharov:2019}, but
has never been demonstrated in the realistic calculations of the $^{8}$He g.s.\
WF.

It should be understood that the strongly correlated source shown in Fig.\
\ref{fig:shulg-sour} (b) is related to about $30 \%$ of the $^{8}$He g.s.\ WF.
The other components of these WF projected on $^{4}$n system populate
``excited'' configurations of $^{4}$n, where we do not expect any strong
FSI effect.

\begin{figure}
\begin{center}
\includegraphics[width=0.47\textwidth]{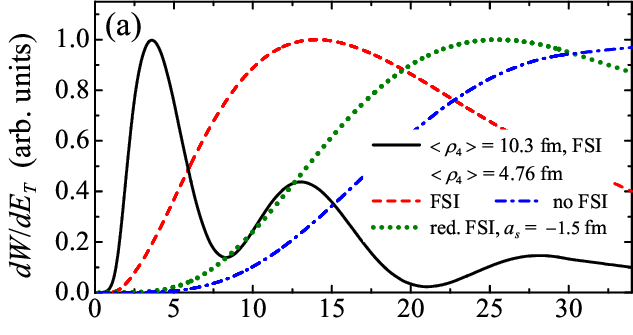}
\includegraphics[width=0.47\textwidth]{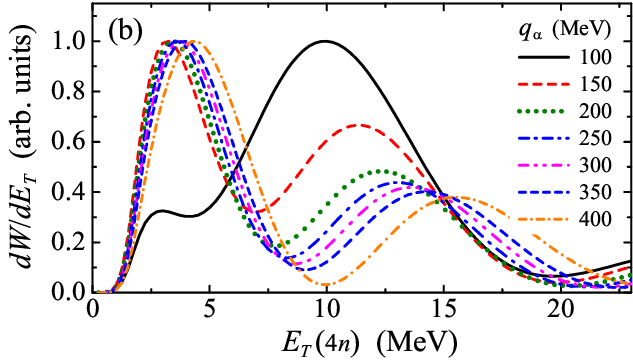}
\end{center}
\caption{(a) The $^{4}$n spectra obtained for $q_{\alpha}=250$ MeV without FSI,
with reduced $n$-$n$ FSI ($a_s=-1.5$  fm), full $^{4}$n FSI with normal source
(colored curves, $\langle \rho_4 \rangle = 4.76$ fm) and with peripheral source 
(black solid curve, $\langle \rho_4 \rangle = 10.3$ fm). (b) $^{4}$n spectra 
obtained for different transferred momenta $q_{\alpha}$ for peripheral source 
with $\langle \rho_4 \rangle = 10.3$ fm.}
\label{fig:shulg-spec}
\end{figure}

Figure \ref{fig:shulg-spec} (a) shows the evolution of $^{4}$n strength
functions for one selected $q_{\alpha}=250$ MeV: no FSI $\rightarrow$ reduced 
FSI
$\rightarrow$ full $^{4}$n FSI. The strength function peak for full $^{4}$n FSI
is at about 14 MeV in a good agreement with analogous calculation in
\cite{Grigorenko:2004}. To form extreme low-energy peak in the strength function
in the region $2.5-3.5$ MeV as indicated by data, a further
qualitative assumption of the peripheral character of reaction is needed. To
evaluate it, we cut out small hyperradii of the original source function by a
profile function of the form
\begin{equation}
f_{\mathbf{q},K\gamma}(\rho_4) \; \rightarrow \;
\frac{f_{\mathbf{q},K\gamma}(\rho_4)}{
 1+\exp[(\rho_{4f}-\rho_4)/d_{4f}]}  \,,
\label{eq:s-f-cut}
\end{equation}
and further refer it as ``peripheral source''. With $\rho_{4f}=10$ fm and
$d_{4f}=0.3$ fm the rms size of the source is increased to $\langle \rho_4
\rangle = 10.3$ fm and the desired energy position of the peak is obtained. It
should be noted that the low-energy peak in the latter calculation is associated
with ``cross'' component of the source, while the higher-energy peak at about
$13$ MeV can be related to the ``broom'' correlation.

The systematics of the strength functions for peripheral sources for different
$q_{\alpha}$ momentum transfers is illustrated in Fig.\
\ref{fig:shulg-spec} (b). One may see that even the ``peripheral assumption'' is
not sufficient to form a strong and distinct low-energy peak in the $^{4}$n
spectrum: the desired picture is realized only for certain intermediate range
of transferred momenta $q_{\alpha}\gtrsim 100$ MeV. This range well corresponds
to situation of our experiment, see Fig.\ \ref{fig:mom-trans} and the related
discussion in Section \ref{sec:popul}.

The calculations for $^{4}$n are performed up to $K_{\max}=14$, which is
insufficient for complete energy convergence. For that reason we perform
simultaneously the calculations with advanced AV14 potential and simple central
$n$-$n$ potential BJ with much faster and ``safer'' convergence. The convergence
curves are used for exponential extrapolation to infinite basis. For example,
for calculations of Fig.\ \ref{fig:shulg-spec} (a), the convergence point was
$3.6 \pm 0.12$ MeV, while ``underbinding'' was $\sim 1.4$ MeV for AV14 and $\sim
0.45$ MeV for BJ potentials. For Fig.\ \ref{fig:lazaus} the corresponding values
are $2.7 \pm 0.15$, $\sim 1.8$ and $\sim 0.7$ MeV. The additional control of the
energy may be provided by phenomenological few-body potential with
Fermi-function profile
\begin{equation}
V_{A} (\rho) = \frac{V_{A0}}{1+\exp[(\rho-\rho_{A0})/d_{A0}]} \, .
\label{eq:pot-coll}
\end{equation}
The quite ``soft'' 4-body potential (\ref{eq:pot-coll}) with parameters
$\rho_{40}=12$ fm and $d_{40}=3$ fm was used to get the peaks at $E_T(4n)$
energies  pointed by the exponential extrapolation. ``Safety'' of the procedure
in the sense of strength function shape was checked by
$K_{\max}=4 \,\rightarrow \, K_{\max}=14$ extrapolations.

\begin{figure}
\begin{center}
\includegraphics[width=0.48\textwidth]{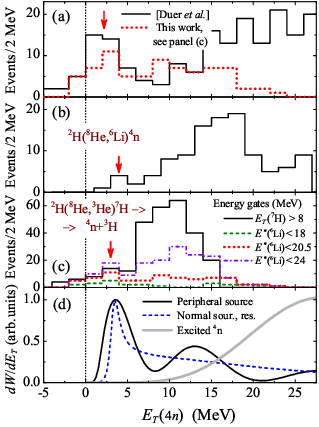}
\end{center}
\caption{Qualitative comparison of theory and data. (a) Data Duer \textit{et
al.} \cite{Duer:2022} plotted together with one of spectra panel (c). (b) Our
data for the $^2$H($^8$He,$^6$Li)$^4$n reaction. (c) Our data for the
$^2$H($^8$He,$^3$He)$^7$H$\rightarrow\,^3$H+$^4$n reaction for different
selection conditions. (d) Theoretical results for $q_{\alpha}=250$ MeV:
nonresonant with peripheral source, resonant for normal source.}
\label{fig:th-comp}
\end{figure}

Qualitative comparison of theory and available data is provided in Fig.\
\ref{fig:th-comp}. Here we need to recall that strongly correlated picture of
Fig.\ \ref{fig:shulg-spec} is connected with $\sim 30 \%$ of the $^{8}$He WF
projecting on the $0^+$ state of $^{4}$n, while $\sim 70 \%$ of this WF populate
excited configurations of the $^{4}$n. The latter WF components are evaluated to
provide strength function with maximum at $30-40$ MeV, see thick gray curve in
Fig.\ \ref{fig:th-comp} (d). As we have mentioned in Section \ref{sec:popul},
the information on transferred momenta is not available in Ref.\
\cite{Duer:2022}, only the rough upper limit estimate $q_{\alpha}<250$ MeV. This
means that, so far, we can not find out even on the
qualitative level, whether the experimental results of this work should be in
principle, the same as in \cite{Duer:2022} (and they just seem somewhat
different only because of experimental resolutions and statistics), or there
should be some significant \emph{physical} difference between them due to
difference in the reaction mechanisms connected with $q_{\alpha}$ value.

It is also important to note that the low-energy peak connected with extreme
peripheral source has very different profile compared to the case of the real
resonant behavior connected with normal-size source. The ``resonant'' (blue
dashed) curve in Fig.\ \ref{fig:th-comp} (d) was generated by additional
artificial binding introduced in Hamiltonian by phenomenological 4-body
potential (\ref{eq:pot-coll}) with parameters $\rho_{40}=7$ fm and $d_{40}=1$
fm. The difference between ``peripheral'' black solid and ``resonant'' blue
dashed curves in Fig.\ \ref{fig:th-comp} (d) is too small to be seriously
discussed for the available quality of the data. However, in principle, we have
guidelines from theory how to distinguish experimentally the ``peripheral''
(caused by reaction mechanism) and ``true resonant'' tetraneutron continuum
responses.

%
%


\section{Theoretical discussion}

\begin{figure}
\begin{center}
\includegraphics[width=0.45\textwidth]{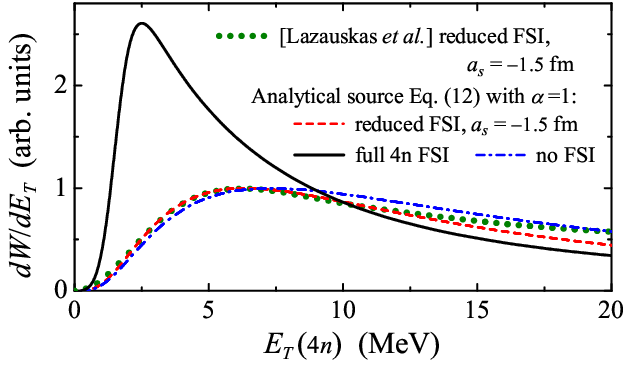}
\end{center}
\caption{Computations simulating the results of Lazauskas \textit{et al.}
\cite{Lazauskas:2023}. Green dotted curve shows calculation with reduced $n$-$n$
interaction from \cite{Lazauskas:2023}. }
\label{fig:lazaus}
\end{figure}

In the recent paper \cite{Lazauskas:2023} the data of \cite{Duer:2022} was
qualitatively reproduced by emission off $^{8}$He-induced source without any
peripheral assumptions. This is in a strong contradiction to the results of
\cite{Grigorenko:2004} and this work. Let's try to track the origin of this
contradiction. In Fig.\ 3 of Ref.\ \cite{Lazauskas:2023} a calculation with
reduced $n$-$n$ interaction (scattering length $a_s=-1.5$ fm) is shown.
This calculation is reasonably close to ``no FSI'' situation and thus its origin
 could be to easier to ``trace''. It is shown in Fig.\ \ref{fig:lazaus} that a
very close result is obtained in our calculations with simple analytical source
\begin{equation}
f_{\mathbf{q},K\gamma}(\rho_4)= \rho^{\alpha}_4 \exp[- \rho_4/\rho_{40}] \,,
\quad \alpha=1 \,, \quad
\rho_{40}=3.85\,\text{fm}\,,
\label{eq:sim-sour}
\end{equation}
for the $K=2$, $S=0$ component. The rms neutron radius $\langle r_n
\rangle=3.34$ fm of such a source is very consistent with the values found in
Table 3 of Ref.\ \cite{Lazauskas:2023}.  The strength function peak shape is
somewhat different from that in Fig.\ 3 of \cite{Lazauskas:2023}, which
indicates more complicated correlated nature of the source function in Ref.\
\cite{Lazauskas:2023}, then we assume in Eq.\ (\ref{eq:sim-sour}). However, the
peak value for the full-calculation $^{4}$n strength function in Fig.\
\ref{fig:lazaus} well coincide with 2.5 MeV of \cite{Lazauskas:2023}, which
means that we are reasonably ``on a right track''.

So, in principle, there exist a class of source functions, which can produce 
extreme low-energy peak in the $^{4}$n spectrum for quite a compact ($\langle 
r_n \rangle \sim 3$ fm) source. The only problem is that the ``supersoft'' 
source Eq.\ (\ref{eq:sim-sour}) has short-distance asymptotic $\sim \rho_{4}$. 
If we get back to the definition Eq.\ (\ref{eq:sour-def}), this implies that 
$^{8}$He overlap with 4-body hypersphericals is singular as $\sim \rho^{-3}_{4}$ 
at zero, which is highly unlikely for any realistic $^{8}$He structure. The 
$^{4}$n sources deduced in our work has asymptotic $\sim \rho^6_{4}$ at zero for 
$K=2$; they can be reasonably approximated by expression like 
(\ref{eq:sim-sour}), but only for $\alpha=6$. This is in full agreement with 
analytically deduced source functions of Ref.\ \cite{Grigorenko:2004}. The 
source with $\alpha=6$ is evidently much ``harder'' in the momentum space than 
the source of the same radius with $\alpha=1$. Therefore, an additional 
\emph{peripheral assumption} is needed to get sizable low-energy effect in 
$^{4}$n spectrum with such a source.

So, we guess that the low-energy enhancement of the $^{4}$n spectrum in Ref.\ 
\cite{Lazauskas:2023} is connected with ``supersoft'' $\alpha=1$ source. The 
origin of the ``supersoft'' source in this work is probably due
to the fact that $^{4}$n c.m.\ is assumed to coincide with $^{8}$He c.m.\ in
Ref.\ \cite{Lazauskas:2023} leading to several important inconsistencies:

\noindent (i) This is an implicit assumption of infinitely heavy
$\alpha$-core in $^{8}$He and, of course, quite a poor approximation considering
the actual mass ratio of $\alpha$ and $^{4}$n.

\noindent (ii)  This also means that within the
cluster model the charge radius of $^{8}$He coincides with the charge radius of
the $\alpha$-particle, which contradicts to the experimental data
\cite{Mueller:2007}.

\noindent (iii)  The ``infinitely heavy core''
assumption also leads to total absence of such an important characteristic as
$q_{\alpha}$ in the
formalism of \cite{Lazauskas:2023}. In contrast, the ``geometric'' effects of
$^{4}$n motion relative to $\alpha$ in $^{8}$He WF are found to be very
pronounced in our calculations, see Fig.\ \ref{fig:shulg-sour}.

\noindent (iv)  The rms size of the source $\langle \rho_4 \rangle = 4.8$ fm, 
see e.g.\ Fig.\ \ref{fig:shulg-spec} (a), presume significantly smaller $\langle 
r_n \rangle = 2.4$ fm distances in the $^{4}$n source, than the $\langle r_n 
\rangle = 2.72$ fm distances in the initial $^{8}$He WF, see Table 
\ref{tab:geom}. The ``infinitely heavy core'' assumption leads to equal $\langle 
r_n \rangle$ distances in the  $^{4}$n source and in the initial $^{8}$He WF.

So, we insist that \emph{peripheral assumption} is necessary to describe the
extreme low-energy peak in the spectrum of $^{4}$n. This point is also
qualitatively supported by extreme small cross sections for the observed
peculiarities in the low-energy $^{4}$n spectra. Typical cross section values
for analogous
direct reactions are some \emph{millibarns} and even \emph{tens of millibarns},
while we and Refs.\ \cite{Kisamori:2016,Duer:2022} observe some 
\emph{microbarns} and even \emph{nanobarns}. Such a cross
section suppression looks quite natural if only the extreme periphery of the
original $^{4}$n configuration of $^{8}$He is actually participating in the 
reactions of interest \emph{without destructive rescattering effects}.


\section{Conclusion}


Evidence for a hump in the $^{4}n$ continuum at $3.5 \pm 0.7$ and $3.2 \pm 0.8$ 
MeV was observed in the reactions $^2$H($^8$He,$^6$Li)$^4$n and 
$^2$H($^8$He,$^3$He)$^7$H$\rightarrow ^3$H+$^4$n, respectively. The $^{4}$n peak 
energies in our work and the ``resonance-like structure'' at $2.4 \pm 0.6$ MeV 
observed in Ref.\ \cite{Duer:2022} are consistent within experimental 
resolutions. We demonstrate that the reaction mechanism in both reported here 
reactions is actually best interpreted as transfer reaction to the ground state 
and to the $E^{\ast}(^{6}\text{Li})=18$ MeV excited state of $^{6}$Li, 
respectively. The reaction mechanism also can be found reasonably consistent 
with that in the $^1$H($^8$He,$p$\,$^4$He)$^4$n reaction \cite{Duer:2022}, so it 
could be the same type of phenomenon. The statistics of 6 events for the first 
our reaction is quite low, but $26-40$ events for the second one is comparable 
to statistics connected with the low-energy $^4$n resonance-like structure 
observed in \cite{Duer:2022}.

Paper \cite{Lazauskas:2023} is explaining the low-energy tetraneutron peak in
Ref.\ \cite{Duer:2022} as solely an initial state ($^{8}$He structure) effect in
the presence of $^{4}$n FSI. This is in a strong contradiction to the results of
\cite{Grigorenko:2004} and the theoretical results of this work. As compared to
\cite{Grigorenko:2004} the calculations of this work (i) are performed with
realistic $NN$ and $3N$ interactions, and (ii) the obtained realistic $^{8}$He
source demonstrates the complicated correlated behavior that is very important
for the $^{4}$n population. So far, the strong spatial correlations (Pauli
focusing) for the $^{8}$He g.s.\ WF has been several times discussed in the
qualitative models \cite{Zhukov:1994,Mei:2012,Sharov:2019}, but in our work they
are for the first time explicitly demonstrated in the realistic calculations of
$^{8}$He g.s.\ structure.

The two aspects mentioned above are not sufficient to fully explain the presence
of a low-lying peak in the tetraneutron spectrum. To do this, it is also
necessary to consider extremely peripheral reaction mechanism, in which the
$\alpha$-core of $^{8}$He interacts with the target, but four neutrons remain
``spectators'' at distances, far exceeding the typical sizes of neutron orbitals
in $^{8}$He. Using the appropriate profile function of the type Eq.\
(\ref{eq:s-f-cut}), simulating such a geometry, it is possible to explain the
presence of a low-lying peak in the tetraneutron spectrum. Since the appearance
of the low-energy peak is not related to the tetraneutron \emph{per se}, but
also to the ISS (the source used) and the reaction mechanism, it should be
expected that such a structure could be detected at somewhat different energies
in different reactions. It is demonstrated that a likely reason of disagreement 
with the calculations
\cite{Lazauskas:2023} may be the incorrect $^{4}$n c.m.\ treatment in the
$^{8}$He WF used in \cite{Lazauskas:2023}.

The existence of low-energy tetraneutron \emph{resonance} would mean a radical
revision of everything we know about neutron-rich nuclei and neutron matter. Our
vision of the problem is that a solution can be found, which is much less
radical being related to the $^{8}$He structure and reaction mechanisms. We got
an important indication in our data that the $^{4}$n spectrum is strongly
dependent on the particular state of the recoil $^{6}$Li system, which means,
first of all, \textit{on the reaction mechanism}. If our understanding of the
low-energy structures in tetraneutron is true, then it is actually not
disappointing at all not to find tetraneutron \emph{resonance}: we get here an
important cautionary lesson concerning low-energy observables in exotic nuclei
and (for those interested) we arrive at a new exciting field of studies of
extreme-peripheral phenomena in nuclear systems.

It is clear that the further studies of $^{8}$He induced reactions on the
deuteron target are needed to bring to full reliability the evidences obtained
in this work. However, the processes of the major interest are all in the cross
sections range of microbarns. Therefore, such studies make sense only if the
statistical limitations of the existing data are overcome by about an order of
magnitude, which is a real challenge for available experimental techniques
and facilities.

%
\textit{Acknowledgments}
%
%
--- This work was supported in part by the Russian Science Foundation (grant
No.\ 22-12-00054). The research was supported in part in the framework of
scientific program of the Russian National Center for Physics and Mathematics,
topic number 6 ``Nuclear and radiation physics'' (2023-2025 stage). The authors
are grateful to Prof.\ Yu.Ts.\ Oganessian for the long-term support of this
research.


\bibliographystyle{apsrev4-2}
\bibliography{d:/latex/all}


\end{document}